\documentclass[conference]{IEEEtran}
\IEEEoverridecommandlockouts

\usepackage[T1]{fontenc}
\usepackage{cite}
\usepackage{amsmath,amssymb,amsfonts}
\usepackage{graphicx}
\usepackage{textcomp}
\usepackage{multirow}
\usepackage{url}

\def\BibTeX{{\rm B\kern-.05em{\sc i\kern-.025em b}\kern-.08em
    T\kern-.1667em\lower.7ex\hbox{E}\kern-.125emX}}

\begin{document}

\title{TMF-RSE: Tri-Modal Fusion with Regional Semantics and Evidential Uncertainty for Lung Severity Scoring}

\author{
\IEEEauthorblockN{Fadi Abdeladhim Zidi\IEEEauthorrefmark{1},
Salah Eddine Bekhouche\IEEEauthorrefmark{2},
Abdellah Zakaria Sellam\IEEEauthorrefmark{1}\IEEEauthorrefmark{3}\\
Gaby Maroun\IEEEauthorrefmark{2},
Fadi Dornaika\IEEEauthorrefmark{2},
Cosimo Distante\IEEEauthorrefmark{1}\IEEEauthorrefmark{3}}
\IEEEauthorblockA{\IEEEauthorrefmark{1}Institute of Applied Sciences and Intelligent Systems (ISASI), CNR, 73100 Lecce, Italy\\
\IEEEauthorrefmark{2}Department of Computer Science and Artificial Intelligence, University of the Basque Country (UPV/EHU), Spain\\
\IEEEauthorrefmark{3}University of Salento, 73100 Lecce, Italy}
}

\maketitle

\begin{abstract}
Accurate quantification of lung disease severity from chest imaging is critical for clinical decision-making and resource allocation. We propose a tri-modal deep learning framework, \textbf{TMF-RSE} (Tri-Modal Fusion with Regional Semantics and Evidential Uncertainty), that combines appearance features from two-dimensional chest inputs, structural features from lung segmentation masks, and semantic features from vision-language models (VLMs) for severity quantification. Our approach employs complementary fusion mechanisms that integrate semantic guidance, structural priors, and hierarchical interactions across modalities. The model employs evidential regression to provide both severity predictions and uncertainty estimates. Experiments on the Per-COVID-19 CT and RALO datasets show that TMF-RSE outperforms recent transformer-based baselines, achieving MAE of 4.02 and Pearson correlation of 0.9629 on Per-COVID-19 validation, and 0.339 MAE / 0.973 PC on RALO geographic extent.
\end{abstract}

\begin{IEEEkeywords}
Lung disease severity quantification, multi-modal fusion, vision-language models, evidential learning, chest CT, chest radiography.
\end{IEEEkeywords}

\section{Introduction}
\label{sec:introduction}

Accurate and automated severity assessment from chest imaging is critical for clinical decision-making, treatment planning, and resource allocation. Continuous severity quantification better reflects gradual parenchymal change than binary or categorical diagnosis, enabling personalized treatment strategies and longitudinal outcome monitoring. Established radiographic scoring frameworks the Brixia index~\cite{borghesi2020brixia,borghesi2022mbrixia}, the RALE scheme~\cite{warren2018rale}, and structured finding taxonomies such as CheXpert~\cite{irvin2019chexpert} have underpinned early automated severity estimation on chest imaging. The COVID-19 pandemic further intensified demand for automated regression, motivating the Per-COVID-19 CT suite~\cite{bougourzi2021percovid,bougourzi2024percovid_challenge} providing percentage-style CT involvement targets under a validation phase with radiologist-estimated references and a stricter test phase with segmentation-derived references~\cite{bougourzi2024percovid_challenge} and the RALO chest radiograph benchmark~\cite{cohen2021ralo_dataset}, defining continuous \emph{geographic extent} and \emph{lung opacity} targets adopted by the transformer regressors~\cite{slika2025lung,slika2025pvitgattip} that serve as our direct baselines. Manual scoring across both modalities remains time-consuming, subjective, and prone to inter-observer variability, motivating fully automated pipelines.

Despite substantial progress from vision transformer and pyramid architectures including ViTReg-IP~\cite{slika2024vitregip}, MViTReg-IP~\cite{slika2025multi}, QCross-Att-PVT~\cite{slika2025lung}, and PViTGAtt-IP~\cite{slika2025pvitgattip} existing methods predominantly exploit appearance cues without incorporating explicit structural priors from lung segmentation or clinical semantics from medical vision--language models (VLMs). While image--mask fusion is increasingly common~\cite{chen2021multimodal}, it typically reduces to concatenation or shallow gating rather than hierarchical semantic conditioning. On the language-vision side, models such as LLaVA-Med~\cite{li2023llava}, BiomedCLIP~\cite{zhang2023biomedclip}, MedCLIP~\cite{wang2022medclip}, and MedGemma~\cite{sellergren2025medgemma} have substantially advanced biomedical image--text understanding, yet remain largely unapplied to calibrated continuous severity regression with mask-aware fusion. Concurrently, evidential regression~\cite{amin2020evidential,gao2024evidential_survey} offers principled separation of aleatoric and epistemic uncertainty, yet few chest severity systems combine evidential outputs with tri-modal fusion and representation-level structural confidence a gap this work directly addresses.

We propose \textbf{TMF-RSE} (Tri-Modal Fusion with Regional Semantics and Evidential Uncertainty), unifying three complementary information sources under a single end-to-end architecture: (1)~\emph{appearance features} from two-dimensional chest inputs capturing intensity and parenchymal patterns; (2)~\emph{structural features} from SAM-derived segmentation masks~\cite{chen2024sam3} encoding anatomical boundaries and spatial relationships; and (3)~\emph{semantic features} from a frozen LLaVA-Med encoder~\cite{li2023llava} applied independently to upper, middle, and lower lung zones via region-specific clinical prompts. Targeting the same Per-COVID-19 CT and RALO protocols as~\cite{slika2025lung,slika2025pvitgattip}, our key contributions are:

\begin{itemize}
    \item A tri-modal fusion architecture with semantic gating, structural prior modulation, and hierarchical fusion capturing joint cross-modal interactions beyond what pairwise combinations can express.
    \item Regional semantic encoding via frozen VLM embeddings with anatomy-aware clinical prompts for upper, middle, and lower lung zones, enabling spatially-grounded severity assessment.
    \item Evidential regression providing aleatoric and epistemic uncertainty estimates with complementary uncertainty modeling at both representation and output levels.
    \item Modality ablations across image-only, image+mask, image+VLM, and full tri-modal configurations, quantified by MAE, RMSE, and Pearson correlation on both benchmarks.
\end{itemize}

\section{Methodology}
\label{sec:methodology}

\begin{figure*}[t]
\centering
\includegraphics[width=1.0\textwidth]{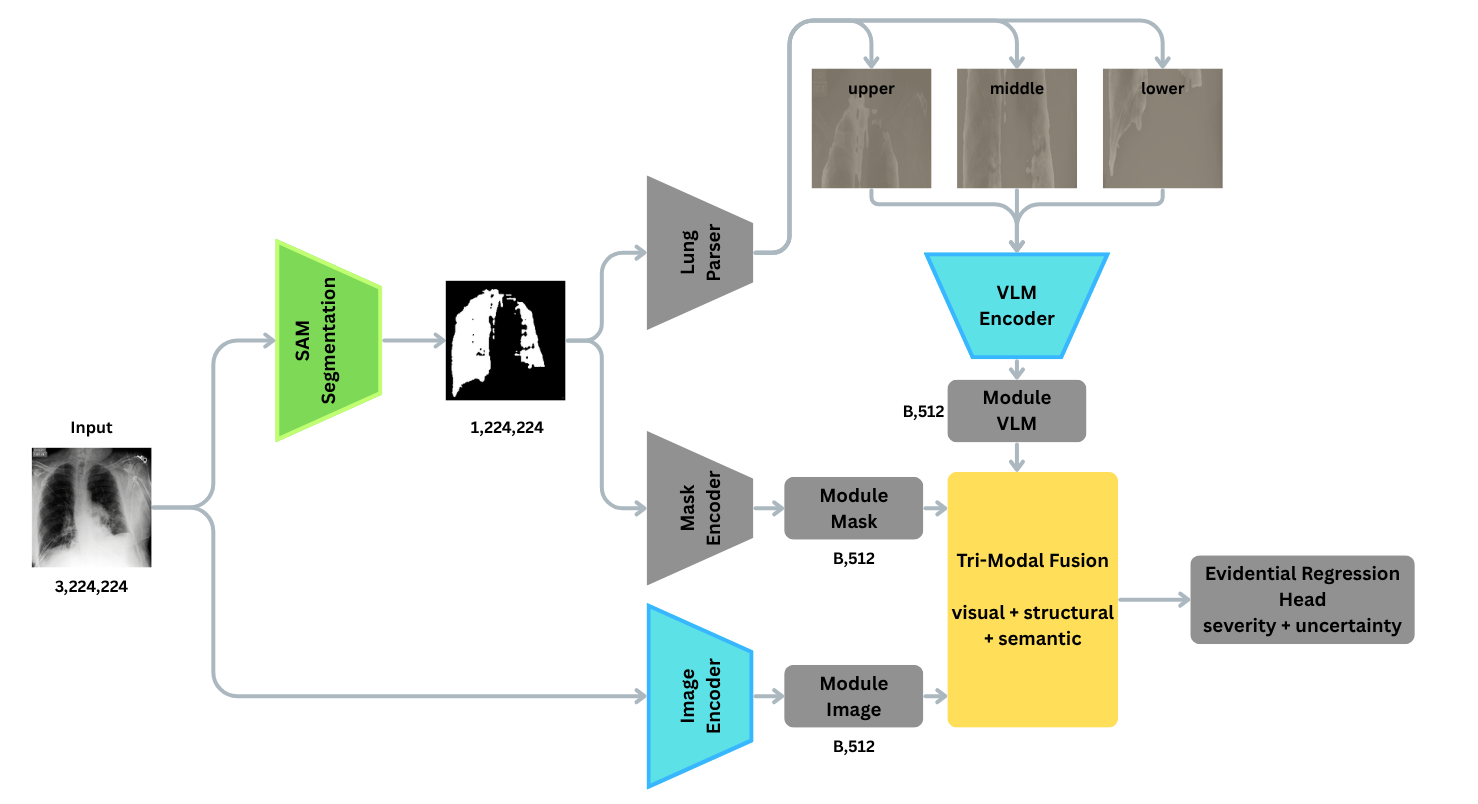}
\caption{Overview of TMF-RSE, a tri-modal framework that integrates appearance, structural, and semantic features for severity quantification with evidential uncertainty estimation.}
\label{fig:architecture}
\end{figure*}

Our tri-modal framework consists of three encoders extracting complementary features, a fusion module that combines them through complementary mechanisms, and an evidential regression head for severity prediction with uncertainty quantification. The overall architecture is illustrated in Figure \ref{fig:architecture}

\subsection{Architecture Overview}


Our proposed architecture (see Fig.~\ref{fig:architecture}) establishes a comprehensive pipeline that processes each two-dimensional chest input whether a CT slice or radiograph through three synergistic encoding pathways. First, the \textbf{Image Encoder} leverages a DINOv3-ViT backbone \cite{oquab2023dinov3} to extract high-fidelity appearance features, effectively capturing nuanced intensity patterns and parenchymal findings. Concurrently, the \textbf{Mask Encoder} utilizes a lightweight Convolutional Neural Network (CNN) to process binary lung segmentation masks generated by SAM3 \cite{chen2024sam3}, thereby explicitly encoding critical anatomical boundaries and structural priors. To contextualize these visual cues, the \textbf{VLM Encoder} employs LLaVA-Med \cite{li2023llava} to distill semantic features from the upper, middle, and lower lung regions via targeted clinical prompts. Ultimately, these complementary representations are harmonized within our tri-modal fusion module and subsequently passed to an evidential regression head, which jointly outputs continuous severity scores and calibrated uncertainty estimates.

\subsection{Image Encoder}

We employ DINOv3-ViT-S/16 \cite{oquab2023dinov3}, a self-supervised vision transformer pretrained on ImageNet, to extract appearance features. DINOv3 provides high-quality dense features that capture fine-grained visual patterns without requiring medical image pretraining. The encoder outputs 384-dimensional features that are projected to 512 dimensions for consistency with other modalities.

\subsection{Mask Encoder}

The mask encoder is a lightweight CNN that processes binary lung segmentation masks to extract structural features. The architecture consists of three convolutional blocks with GroupNorm and ReLU activations, followed by global average pooling and a projection layer. This encoder captures anatomical boundaries, lung shape, and spatial relationships that complement appearance features.

\subsection{VLM Encoder with Regional Semantics}

We use LLaVA-Med v1.5-Mistral-7B \cite{li2023llava}, a medical vision-language model, to extract semantic features from lung regions. Importantly, the VLM is not used for direct prediction but as a semantic prior that regularizes visual reasoning. Freezing the VLM ensures stable clinical semantics while preventing overfitting and excessive computational cost during training. The lungs are divided into three vertical zones (upper, middle, lower), adopting a simple three-zone division to balance anatomical relevance and robustness, avoiding reliance on fine-grained anatomical landmarks that may be unreliable in severe disease. Each region is processed separately with region-specific clinical prompts:

To extract anatomically grounded semantic features, we employ region-specific clinical prompts tailored to the lung's spatial architecture. Specifically, the VLM is queried to characterize \textbf{upper} field opacities, \textbf{middle} zone ground-glass and interstitial patterns, and \textbf{lower} base consolidation and atelectasis. This targeted prompting ensures the distilled representations reflect localized pathological signatures critical for robust severity assessment.

Each region produces a 768-dimensional semantic embedding, resulting in a $3 \times 768$ feature tensor that captures spatially-aware clinical semantics.

\subsection{Tri-Modal Fusion Module}

Our fusion module employs three complementary mechanisms that work together to integrate multi-modal information:

\subsubsection{Semantic Gating}
VLM-derived semantic features generate gates that selectively control which image features are used. This prevents the model from over-relying on spurious visual patterns by ensuring only semantically-relevant image evidence is passed forward:

\begin{equation}
\mathbf{g} = \sigma(\text{MLP}(\mathbf{f}_{\text{vlm}})), \quad \mathbf{f}_{\text{img}}^{\text{gated}} = \mathbf{f}_{\text{img}} \odot \mathbf{g}
\end{equation}

where $\mathbf{g} \in [0,1]^d$ are gate values, $\sigma$ is sigmoid, and $\odot$ denotes element-wise multiplication.

\subsubsection{Structural Prior Modulation}
Mask features influence attention entropy, reflecting anatomical confidence. Well-defined lung boundaries produce sharp attention (low entropy), while poor segmentation leads to diffuse attention (high entropy), signaling model uncertainty:

\begin{equation}
c_{\text{struct}} = \sigma(\text{MLP}(\mathbf{f}_{\text{mask}})), \quad \mathbf{a}_{\text{mod}} = c_{\text{struct}} \cdot \mathbf{a} + (1-c_{\text{struct}}) \cdot \mathbf{a}_{\text{uniform}}
\end{equation}

where $c_{\text{struct}}$ is structural confidence and $\mathbf{a}$ are attention weights. Structural prior modulation influences feature uncertainty at the representation level, while evidential regression models predictive uncertainty at the output level, allowing complementary uncertainty modeling across the network.

\subsubsection{Hierarchical Fusion}
We implement a three-stage hierarchical fusion scheme to progressively integrate cross-modal interactions. \textbf{Stage 1} synthesizes image and mask features into a structural appearance representation, which is then combined with VLM embeddings in \textbf{Stage 2} to derive semantically-aware structural features. Finally, \textbf{Stage 3} employs a synergy gate to adaptively modulate the unified representation when the joint interaction of all three modalities is essential.
While pairwise fusion captures partial interactions, the proposed hierarchical design explicitly conditions later representations on all three modalities, encouraging joint dependency. The fusion module also includes cross-attention layers for pairwise interactions between modalities, complementing the hierarchical mechanism.
\subsection{Evidential Regression Head}

We employ evidential regression \cite{amin2020evidential} to predict severity scores along with uncertainty estimates. The head outputs parameters of a Normal-Inverse-Gamma (NIG) distribution: $\gamma$ (predicted mean), $\nu$ (virtual observations), $\alpha$ (shape), and $\beta$ (scale). Severity is predicted as $\gamma$, while uncertainties are computed as:

\begin{align}
\text{Aleatoric} &= \frac{\beta}{\alpha - 1} , \quad
\text{Epistemic} &= \frac{\beta}{\nu(\alpha - 1)}
\end{align}

The loss function combines negative log-likelihood with a regularization term that prevents over-confidence:

\begin{equation}
\mathcal{L} = \text{NLL}(\gamma, \nu, \alpha, \beta; y) + \lambda \cdot \text{Reg}(\nu, \alpha)
\end{equation}

where $y$ is the ground truth severity score and $\lambda$ is a regularization coefficient (set to 0.001).

\section{Experiments}
\label{sec:experiments}

\subsection{Datasets}
\label{sec:datasets}

Experiments were conducted on the RALO chest radiograph benchmark~\cite{cohen2021ralo_dataset} and the Per-COVID-19 CT benchmark~\cite{bougourzi2021percovid,bougourzi2024percovid_challenge}.

\textbf{RALO}~\cite{cohen2021ralo_dataset} comprises 2,373 postero-anterior chest radiographs scored by two expert radiologists at Stony Brook Medicine, split into 1,878 training and 495 test images. Each study is graded on two continuous targets: \emph{Geographic Extent} (GE; per-lung 0--4, summed range 0--8) describing hemithorax involvement by ground-glass opacity or consolidation, and \emph{Lung Opacity} (LO; per-lung 0--4, summed range 0--8) reflecting opacity density. Reference labels are the inter-reader mean, yielding values in $\{0, 0.5, \ldots, 8.0\}$~\cite{cohen2021ralo_dataset,slika2025pvitgattip}. Following prior regression work~\cite{slika2025lung,slika2025pvitgattip}, offline combined lung masking with score replacement is applied on the training set.

\textbf{Per-COVID-19 CT}~\cite{bougourzi2021percovid} comprises 189 CT volumes from RT-PCR-confirmed COVID-19 patients (one scan per patient; aged $\approx$27--70 years), reviewed by two experienced thoracic radiologists~\cite{bougourzi2021percovid}. Each volume spans 40--70 axial slices, with radiologists providing COVID-19 Infection Percentage (CIP) targets expressing the fraction of affected lung parenchyma. Following the international challenge protocol~\cite{bougourzi2024percovid_challenge}, results are reported under two regimes: a validation phase with radiologist-estimated references and a test phase with segmentation-derived references (Tables~\ref{tab:percovid-val}--\ref{tab:percovid-test}).

For both benchmarks, inputs are resized to a common spatial resolution (see Sec.~\ref{sec:implementation_details}). Lung masks are generated using SAM3~\cite{chen2024sam3} with a confidence threshold of 0.2 and cached for efficiency.

\subsection{Evaluation Metrics}
\label{sec:evaluation_metrics}
To comprehensively evaluate the performance of severity quantification models, we employ multiple complementary regression metrics that capture both absolute error and ranking consistency between predicted and reference severity scores.

\subsubsection{Mean Absolute Error (MAE)}
Mean Absolute Error (MAE) measures the average absolute difference between predicted severity scores $\hat{s}_i$ and reference severity scores $s_i$ over $N$ samples:
\begin{equation}
\mathrm{MAE} = \frac{1}{N} \sum_{i=1}^{N} \left| \hat{s}_i - s_i \right|.
\end{equation}
MAE provides an intuitive measure of prediction accuracy and is robust to outliers, making it suitable for clinical severity estimation where extreme errors are undesirable.

\subsubsection{Root Mean Squared Error (RMSE)}
Root Mean Squared Error (RMSE) emphasizes larger prediction errors by squaring the residuals before averaging:
\begin{equation}
\mathrm{RMSE} = \sqrt{ \frac{1}{N} \sum_{i=1}^{N} \left( \hat{s}_i - s_i \right)^2 }.
\end{equation}
RMSE is particularly sensitive to severe misestimations and thus reflects the reliability of the model in high-severity cases.

\subsubsection{Pearson Correlation Coefficient}
The Pearson correlation coefficient ($r$) evaluates the linear relationship between predicted and reference severity scores:
\begin{equation}
r = \frac{\sum_{i=1}^{N} (\hat{s}_i - \overline{\hat{s}})(s_i - \overline{s})}
{\sqrt{\sum_{i=1}^{N} (\hat{s}_i - \overline{\hat{s}})^2}
 \sqrt{\sum_{i=1}^{N} (s_i - \overline{s})^2}},
\end{equation}
where $\overline{\hat{s}}$ and $\overline{s}$ denote the mean predicted and reference severity scores, respectively.
Pearson correlation assesses how well the model preserves the relative ordering of severity values under a linear assumption.

\subsection{Implementation Details}
\label{sec:implementation_details}

\subsubsection{Training Configuration}
All models were trained for 50 epochs using the AdamW optimizer with learning rate $10^{-4}$ and weight decay $10^{-5}$. We used a ReduceLROnPlateau scheduler with factor 0.5 and patience 5 epochs. Gradient clipping was applied with maximum norm 1.0. Pretrained encoders (DINOv3 and LLaVA-Med) were frozen during training to maintain their learned representations while allowing the fusion module and regression head to adapt. Despite the small batch size, training remained stable due to gradient accumulation and frozen backbone encoders.

\subsubsection{Data Preprocessing}
Two-dimensional chest inputs (CT slices for Per-COVID-19 CT; radiographs for RALO) were resized to $224 \times 224$ pixels and normalized using ImageNet statistics. Lung masks were generated using SAM3 \cite{chen2024sam3} with a confidence threshold of 0.2, and masks were cached for efficiency. Regions were extracted by dividing the lung mask into three equal vertical zones.
\subsubsection{Model Architecture Details}
The image encoder uses DINOv3-ViT-S/16 with output dimension 512. The mask encoder is a CNN with output dimension 256. The VLM encoder outputs 768-dimensional features per region (3 regions total). The fusion module projects all modalities to a common dimension of 512. Cross-attention uses 8 heads. The evidential regression head has hidden dimensions [256, 128] with LayerNorm and dropout (0.2).

\subsubsection{Computational Resources}
Training was performed on NVIDIA L4 GPUs with batch size 1 (due to VLM processing requirements). Gradient accumulation was used to stabilize optimization despite small batch sizes. Each ablation variant was trained independently, with parallel execution across multiple GPUs when available. Training time varied from 8--12 hours per model depending on the variant. Our ablations explicitly evaluate whether the performance gain from VLM integration justifies the additional computational cost.


\begin{table}[t]
\centering
\caption{RALO benchmark results: geographic extent and lung opacity scores. \textbf{Bold} indicates best; \underline{underline} indicates second-best.}
\label{tab:ralo}
\begin{tabular}{|l|c|c|c|c|}
\hline
\multirow{2}{*}{Method} & \multicolumn{2}{c|}{Geographic extent} & \multicolumn{2}{c|}{Lung opacity} \\ \cline{2-5} 
                        & MAE $\downarrow$ & PC $\uparrow$ & MAE $\downarrow$ & PC $\uparrow$ \\ \hline
ViTReg-IP \cite{slika2024vitregip} & 0.565 & 0.925 & 0.510 & 0.857 \\ \hline
MViTReg-IP \cite{slika2025multi} & 0.531 & 0.938 & 0.462 & 0.881 \\ \hline
QCross-Att-PVT \cite{slika2025lung} & 0.362 & \underline{0.971} & 0.337 & \underline{0.945} \\ \hline
PViTGAtt-IP \cite{slika2025pvitgattip} & \underline{0.351} & 0.963 & \underline{0.322} & 0.941 \\ \hline
\textbf{TMF-RSE (Ours)} & \textbf{0.339} & \textbf{0.973} & \textbf{0.310} & \textbf{0.960} \\ \hline
\end{tabular}
\end{table}



\begin{table}[t]
\centering
\caption{Per-COVID-19 CT, validation phase: comparison with published methods and challenge leaders. \textbf{Bold} indicates best; \underline{underline} indicates second-best.}
\label{tab:percovid-val}
\begin{tabular}{|l|c|c|c|}
\hline
Method & MAE $\downarrow$ & PC $\uparrow$ & RMSE $\downarrow$ \\ \hline
TAC & 4.48 & 0.9460 & 8.54 \\ \hline
Taiyuan\_university\_lab713 & 4.50 & \underline{0.9490} & \underline{8.09} \\ \hline
EIDOSlab\_Unito & 4.91 & 0.9429 & 8.70 \\ \hline
ausilianapoli94 & 4.95 & 0.9435 & 8.60 \\ \hline
ACVLab & 4.99 & 0.9364 & 9.08 \\ \hline
Baseline & 5.24 & 0.9322 & 9.45 \\ \hline
QCross-Att-PVT \cite{slika2025lung} & 5.42 & 0.9432 & N/A \\ \hline
PViTGAtt-IP \cite{slika2025pvitgattip} & \underline{4.09} & 0.9439 & N/A \\ \hline
\textbf{TMF-RSE (Ours)} & \textbf{4.02} & \textbf{0.9629} & \textbf{7.07} \\ \hline
\end{tabular}
\end{table}

\begin{table}[t]
\centering
\caption{Per-COVID-19 CT, testing phase: comparison with published methods and challenge leaders. \textbf{Bold} indicates best; \underline{underline} indicates second-best.}
\label{tab:percovid-test}
\setlength{\tabcolsep}{5pt}
\begin{tabular}{|l|c|c|c|}
\hline
Method & MAE $\downarrow$ & PC $\uparrow$ & RMSE $\downarrow$ \\ \hline
Taiyuan\_university\_lab713 & \textbf{3.55} & \underline{0.8547} & \underline{7.51} \\ \hline
TAC & \underline{3.64} & 0.8022 & 8.57 \\ \hline
SenticLab.UAIC & 4.61 & 0.7634 & 9.09 \\ \hline
ACVLab & 4.86 & 0.7287 & 10.27 \\ \hline
EIDOSlab\_Unito & 5.02 & 0.7977 & 9.01 \\ \hline
IPLab & 6.53 & 0.7091 & 9.97 \\ \hline
Baseline & 8.57 & 0.6344 & 12.62 \\ \hline
QCross-Att-PVT \cite{slika2025lung} & 4.45 & 0.8094 & N/A \\ \hline
PViTGAtt-IP \cite{slika2025pvitgattip} & \textbf{3.55} & \underline{0.8547} & N/A \\ \hline
\textbf{TMF-RSE (Ours)} & 3.69 & \textbf{0.9122} & \textbf{6.15} \\ \hline
\end{tabular}
\end{table}

\section{Results}
\label{sec:results}

Table~\ref{tab:ralo} compares TMF-RSE against published transformer regressors on RALO geographic extent and lung opacity \cite{slika2024vitregip,slika2025multi,slika2025lung,slika2025pvitgattip}. PViTGAtt-IP achieves the strongest reported numbers on this table, particularly on lung opacity MAE/PC \cite{slika2025pvitgattip}; QCross-Att-PVT is competitive on geographic extent \cite{slika2025lung}. \textbf{TMF-RSE (Ours)} reports our full model under the same protocol.

Table~\ref{tab:percovid-val} and Table~\ref{tab:percovid-test} place the same published methods \cite{slika2025lung,slika2025pvitgattip} next to challenge leaderboard entries \cite{bougourzi2024percovid_challenge}. Pearson correlation remains high on validation (radiologist references) for several teams and for PViTGAtt-IP, but drops markedly on the test phase when segmentation-derived references are used consistent with the greater difficulty of that regime \cite{bougourzi2024percovid_challenge}. Leaderboard MAE/RMSE on test (e.g., Taiyuan\_university\_lab713 vs.\ TAC) should be interpreted jointly with PC, because ranking can differ by metric.

\begin{table}[t]
    \centering
    \caption{Modality ablation: image only, pairwise bi-modal variants, and full TMF-RSE. \textbf{Bold} indicates the best performance; \underline{underline} indicates the second-best.}
    \label{tab:ablation}
    \footnotesize
    \setlength{\tabcolsep}{3pt}
    \begin{tabular}{|l|cc|cc|cc|}
    \hline
    \multirow{2}{*}{Variant} & \multicolumn{2}{c|}{RALO: GE} & \multicolumn{2}{c|}{RALO: LO} & \multicolumn{2}{c|}{Per-COVID-19 CT (val)} \\
    \cline{2-7}
     & MAE$\downarrow$ & PC$\uparrow$ & MAE$\downarrow$ & PC$\uparrow$ & MAE$\downarrow$ & PC$\uparrow$ \\
    \hline
    Image only & 0.469 & 0.9568 & 0.458 & 0.9209 & 4.82 & 0.9436 \\ \hline
    Image + mask & 0.432 & 0.9612 & 0.410 & 0.9347 & 4.67 & 0.9498 \\ \hline
    Image + VLM & \underline{0.395} & \underline{0.9669} & \underline{0.376} & \underline{0.9412} & \underline{4.53} & \underline{0.9517} \\ \hline
    \textbf{TMF-RSE} & \textbf{0.339} & \textbf{0.973} & \textbf{0.310} & \textbf{0.960} & \textbf{4.02} & \textbf{0.9629} \\ \hline
    \end{tabular}
\end{table}

Table~\ref{tab:ablation} summarizes modality ablations under the same evaluation protocol as the main benchmarks (Per-COVID-19 CT: validation phase, as in Table~\ref{tab:percovid-val}).

\subsection{Uncertainty-based sample removal}
\label{sec:uncertainty_sparsification}

To verify that evidential uncertainty identifies difficult cases, we perform an \emph{uncertainty sparsification} analysis on the test set. For each benchmark, samples are sorted by descending total predictive uncertainty (epistemic plus aleatoric from the NIG head; Sec.~\ref{sec:methodology}). We then remove the top fraction of highest-uncertainty samples and recompute mean absolute error (MAE) on the remaining subset. The horizontal axis reports the percentage of removed samples; the vertical axis reports MAE. The curve should trend downward as hard cases are discarded, indicating lower average error on the retained, lower-uncertainty cases.

\begin{figure}[t]
\centering
\includegraphics[width=\columnwidth]{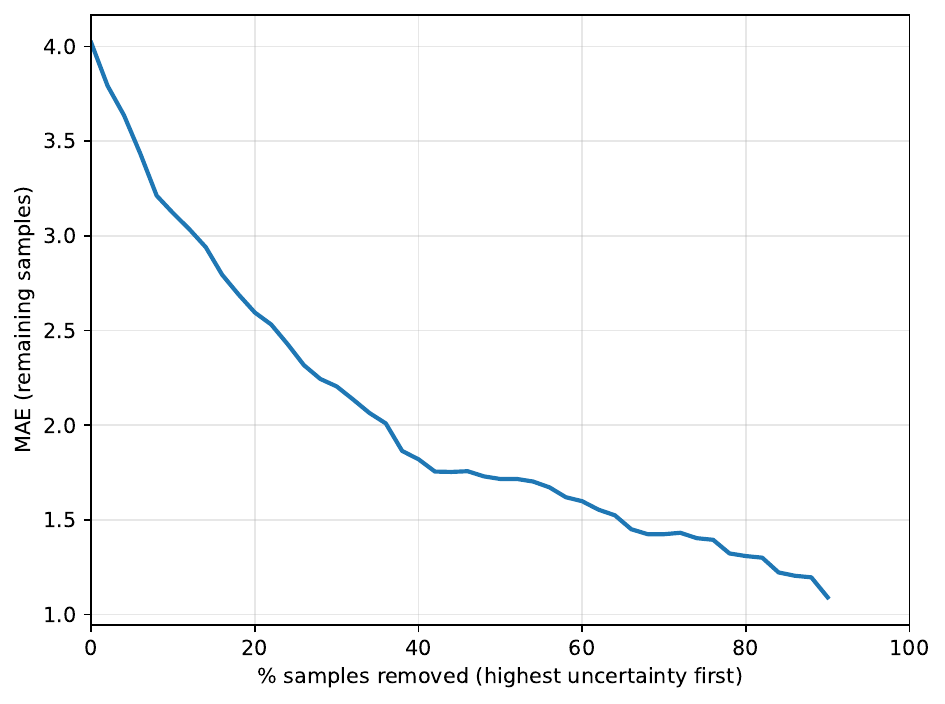}
\caption{MAE as a function of removing high-uncertainty samples. Decreasing MAE confirms that uncertainty identifies difficult cases.}
\label{fig:uncertainty_sparsification}
\end{figure}

Figure~\ref{fig:uncertainty_sparsification} summarizes the expected behavior: progressively removing high-uncertainty predictions reduces MAE on the remainder, supporting the use of evidential scores as case-level difficulty indicators.

\section{Discussion}
\label{sec:discussion}

\subsection{Modality Contributions}

Our ablation studies evaluate the contribution of each modality to severity quantification. Image features provide the foundation, capturing radiographic findings directly. Mask features add structural context, helping the model focus on lung regions and understand anatomical boundaries. VLM features incorporate clinical semantics that guide interpretation of visual patterns through semantic gating mechanisms.

Table~\ref{tab:ablation} shows consistent gains when adding mask structure and VLM semantics; the full model achieves the strongest reported MAE/PC on each column relative to the listed variants. Hierarchical fusion then combines all three pathways jointly (Sec.~\ref{sec:methodology}).

\subsection{Fusion Mechanism Effectiveness}
The architecture pairs hierarchical fusion with cross-attention (Sec.~\ref{sec:methodology}). Semantic gating steers which image evidence is emphasized using VLM-derived features; structural prior modulation links attention behavior to mask-derived confidence. We do not tabulate a concatenation-only baseline here; such a comparison is left for supplementary material if needed.
The hierarchical fusion design conditions later representations on all three modalities, encouraging joint dependency. This suggests that tri-modal fusion captures interactions that are not fully modeled by pairwise combinations.

\subsection{Regional Semantics}
Multi-region prompts (Sec.~\ref{sec:methodology}) target heterogeneity across lung zones. Table~\ref{tab:ablation} reports modality subsets rather than single-region vs.\ multi-region VLM prompts; the latter comparison can be added to supplementary material if available.
\subsection{Uncertainty Quantification}
Evidential regression provides clinically valuable uncertainty estimates alongside severity predictions. Epistemic uncertainty reflects model confidence (lower values indicating higher confidence), while aleatoric uncertainty captures inherent variability in severity labels. These estimates can guide clinical decision-making by flagging cases where predictions are uncertain, enabling clinicians to exercise appropriate caution.
The sparsification curve in Fig.~\ref{fig:uncertainty_sparsification} complements aggregate metrics: removing the highest-uncertainty cases should reduce MAE on the remainder if uncertainty tracks predictive error.
The complementary uncertainty modeling structural prior modulation at the representation level and evidential regression at the output level distinguishes our approach from standard regression methods. This dual-level uncertainty modeling is particularly valuable for clinical applications where confidence in predictions is crucial for decision-making.
\subsection{Limitations and Future Work}
Our study has several limitations. First, we evaluate on the Per-COVID-19 CT and RALO datasets; validation on additional cohorts and modalities would strengthen generalizability. Second, the VLM encoder is computationally expensive, limiting batch size and training efficiency. Future work could explore more efficient VLM architectures or knowledge distillation to reduce computational cost.
Third, although Tables~\ref{tab:ralo}--\ref{tab:percovid-test} include strong published baselines \cite{slika2025lung,slika2025pvitgattip}, external validation on held-out hospitals, scanners, and non-COVID etiologies is still needed to establish clinical transportability beyond these benchmarks.
Fourth, benchmark-specific label definitions (radiologist estimates vs.\ segmentation-derived scores on Per-COVID-19 CT; continuous opacity and extent targets on RALO) may not capture all clinical nuances. Future work could explore learned severity objectives or incorporate additional clinical factors.
Finally, clinical validation is needed to assess whether uncertainty estimates correlate with clinical outcomes and whether the model's predictions improve clinical decision-making in practice. The interpretability provided by semantic gating and attention mechanisms should facilitate such validation.
\section{Conclusion}
\label{sec:conclusion}

We presented a tri-modal framework for lung disease severity quantification that combines appearance, structural, and semantic features. Our approach employs three complementary mechanisms semantic gating, structural prior modulation, and hierarchical fusion designed to leverage interactions across all three modalities.
Table~\ref{tab:ablation} reports modality ablations on RALO and Per-COVID-19 CT (validation): image-only, image+mask, image+VLM, and full TMF-RSE. Evidential outputs and fusion design are described in Sec.~\ref{sec:methodology}; MAE under uncertainty-based sample removal is summarized in Sec.~\ref{sec:uncertainty_sparsification} and Fig.~\ref{fig:uncertainty_sparsification}, with broader calibration studies left for future work. 
The model provides uncertainty estimates through evidential regression, offering both aleatoric and epistemic uncertainty that can guide clinical decision-making. The design enables complementary uncertainty modeling at both representation and prediction levels, distinguishing it from standard regression approaches.

Future work will focus on improving computational efficiency, validating on additional datasets, and conducting clinical studies to assess real-world impact. The tri-modal framework provides a foundation for future research in medical image analysis that seeks to combine visual, structural, and semantic understanding.

\bibliographystyle{IEEEtran}
\bibliography{references}

\end{document}